\documentclass[sigconf,screen,authorversion]{acmart}

\AtBeginDocument{%
  }
\setcopyright{acmcopyright}
\copyrightyear{2023}
\acmYear{2023}
\acmDOI{XXXXXXX.XXXXXXX}
\acmConference[]{}{}{}

\usepackage[normalem]{ulem}
\usepackage{comment}
\usepackage{booktabs}
\usepackage{multicol}
\usepackage{multirow}
\usepackage{prettyref}
\usepackage{hyperref}
\newrefformat{eqn}{Equation~\ref{#1}}
\newrefformat{fig}{Figure~\ref{#1}}
\newrefformat{tab}{Table~\ref{#1}}
\newrefformat{sec}{Section~\ref{#1}}
\newrefformat{ssec}{Section~\ref{#1}}
\newrefformat{lst}{Listing~\ref{#1}}

\usepackage{caption}
\usepackage{subcaption}
\usepackage{graphicx}
\graphicspath{ {./pictures/} }

\usepackage{color}
\usepackage{soul}
\usepackage{siunitx}

\newcommand{\ligen}{LiGen}

\begin{document}

\title{Improving computation efficiency using input and architecture features for a virtual screening application}

\author{Accordi Gianmarco}
\orcid{0000-0001-8023-2108}
\affiliation{%
  \institution{Dipartimento di Elettronica, Informazione e Bionigegneria Politecnico di Milano}
  \streetaddress{Via Giuseppe Ponzio, 34}
  \city{Milano}
  \country{Italy}
}
\email{gianmarco.accordi@polimi.it}

\author{Vitali Emanuele}
\orcid{0000-0001-8629-2099}
\affiliation{%
  \institution{CSC – IT Center for Science}
  \streetaddress{P.O. Box 405}
  \city{Espoo}
  \country{Finland}
}
\email{emanuele.vitali@csc.fi}

\author{Gadioli Davide}
\orcid{0000-0002-0143-0737}
\affiliation{%
  \institution{Dipartimento di Elettronica, Informazione e Bionigegneria Politecnico di Milano}
  \streetaddress{Via Giuseppe Ponzio, 34}
  \city{Milano}
  \country{Italy}
}
\email{davide.gadioli@polimi.it}

\author{Crisci Luigi}
\affiliation{%
  \institution{Dipartimento di Informatica
  Università degli studi di Salerno}
  \streetaddress{Via Giovanni Paolo II, 132}
  \city{Salerno}
  \country{Italy}
}
\email{lcrisci@unisa.it}

\author{Cosenza Biagio}
\orcid{0000-0002-8869-6705}
\affiliation{%
  \institution{Dipartimento di Informatica
  Università degli studi di Salerno}
  \streetaddress{Via Giovanni Paolo II, 132}
  \city{Salerno}
  \country{Italy}
}
\email{bcosenza@unisa.it}

\author{Bisson Mauro}
\affiliation{%
  \institution{NVIDIA Corporation}
  \streetaddress{Menlo Park}
  \city{Santa Clara}
  \country{CA, USA}
}
\email{maurob@nvidia.com}

\author{Massimiliano Fatica}
\orcid{0000-0002-5839-1644}
\affiliation{%
  \institution{NVIDIA Corporation}
  \streetaddress{Menlo Park}
  \city{Santa Clara}
  \country{CA, USA}
}
\email{mfatica@nvidia.com}

\author{Beccari Andrea}
\orcid{0000-0001-6830-2695}
\affiliation{%
  \institution{Dompé farmaceutici SpA}
  \streetaddress{Via Santa Lucia, 6}
  \city{Napoli}
  \country{Italy}
}
\email{andrea.beccari@dompe.com}

\author{Gianluca Palermo}
\orcid{0000-0001-7955-8012}
\affiliation{%
  \institution{Dipartimento di Elettronica, Informazione e Bionigegneria Politecnico di Milano}
  \streetaddress{Via Giuseppe Ponzio, 34}
  \city{Milano}
  \country{Italy}
}
\email{gianluca.palermo@polimi.it}


\renewcommand{\shortauthors}{Accordi et al.}

\begin{abstract}

Virtual screening is an early stage of the drug discovery process that selects the most promising candidates.
In the urgent computing scenario it is critical to find a solution in a short time frame.
In this paper, we focus on a real-world virtual screening application to evaluate out-of-kernel optimizations, that consider input and architecture features to improve the computation efficiency on GPU.
Experiment results on a modern supercomputer node show that we can almost double the performance.
Moreover, we implemented the optimization using SYCL and it provides a consistent benefit with the CUDA optimization.
A virtual screening campaign can use this gain in performance to increase the number of evaluated candidates, improving the probability of finding a drug. 

\end{abstract}

\ccsdesc[500]{Computing methodologies~Parallel programming languages}
\ccsdesc[500]{Computer systems organization~Single instruction, multiple data}

\keywords{SYCL, CUDA, parallel programming, virtual screening}


\maketitle

\section{Introduction}
\label{sec:introduction}

Drug discovery aims at finding a small molecule that has a beneficial effect against a target disease.
It involves \textit{in-vitro} and \textit{in-vivo} stages that increase its cost and duration, limiting the number of evaluated candidates.
Recent studies demonstrated that we increase the probability of finding a drug by introducing an \textit{in-silico} stage that selects which molecules to test in-vitro,\cite{matter2011application, allegretti2022repurposing}.
In this stage, we use a \textit{virtual screening} software to estimate the interaction strength of a small molecule, named \textit{ligand}, with the target protein.
We can use this value to rank a chemical library of many ligands. We forward only the most promising ones to the following drug discovery stages.

It is possible to create a chemical library by simulating known chemical reactions.
With this method, we can have access to massive chemical space.
Therefore, the chemical library size is limited only by the computation effort available for the virtual screening.
For this reason, supercomputers are the target system for a virtual screening campaign \cite{9817028,10.1177/10943420211001565}.
When we look at the node architectures of the fastest ones, according to the TOP500\footnote{\url{https://www.top500.org/}} list, we can notice how they heavily rely on accelerators for increasing their throughput.
Even if the accelerators are typically GPUs, they belong to different vendors.
Thus, it is not possible to have a single implementation that uses the accelerator's native language.
Urgent computing becomes relevant in this context because we want to reduce a pandemic social and economic impact, so virtual screening software must benefit from all the computation resources available.
In this work we focus on the \ligen{} virtual screening application as a real-world case study \cite{10.48550/arxiv.2209.05069,crisci2022towards}.
In particular, we introduce a data preparation phase that uses architecture and input features to organize the input of CUDA kernels.
This phase aims at increasing the computation efficiency on GPU.
Moreover, we ported this feature for the SYCL kernel implementation.
From the experimental results, this out-of-kernel optimization increase the performance of both implementations.
Even if the results relate to \ligen{}, they might apply also to other applications with a similar design.


In \prettyref{sec:related_works} we describe the virtual screening problem and we describe the \ligen{} implementation, to better define the out-of-kernel optimizations reported in \prettyref{sec:ook_optimization}, and their porting in SYCL 2020 described in \prettyref{sec:sycl_porting}.
\prettyref{sec:experimental_results} report the experimental evaluation of the proposed optimizations.
Finally, \prettyref{sec:conclusion} concludes the paper.
\section{Background and related works}
\label{sec:related_works}

Three main tasks are required to virtual screen a ligand.
Since the ligand molecule is much smaller than a protein, the first task aims at identifying one or more regions of the protein where we would like to dock the ligand. We name each of these regions as \textit{docking site}.
The second task aims at estimating the 3D displacement of the ligand's atoms when it interacts with the target docking site: it \textit{dock}s the ligand in the protein.
Finally, the third task evaluates the interaction strength between the docked ligand with the protein.
Thus, for each docking site, we can rank the input chemical library to help domain experts to find the best candidates to test in-vitro.

The virtual screening is a well-known problem in literature, where many approaches have been proposed, implemented, and evaluated \cite{Pagadala2017,Yuriev2015}.
In the urgent computing context, there is an effort for re-designing existing software for scaling out and using accelerators of modern heterogeneous HPC nodes \cite{9817028,10.1177/10943420211001565}, reaching a considerable performance improvement.
For example, AutoDock-Vina \cite{tang2022accelerating} and PLANTS \cite{Korb2011} observed a speedup of over 50x after their porting in CUDA. The data parallelism available in the previously cases makes a good match for GPU's architecture.

\begin{figure}[t]
    \includegraphics[width=0.7\columnwidth]{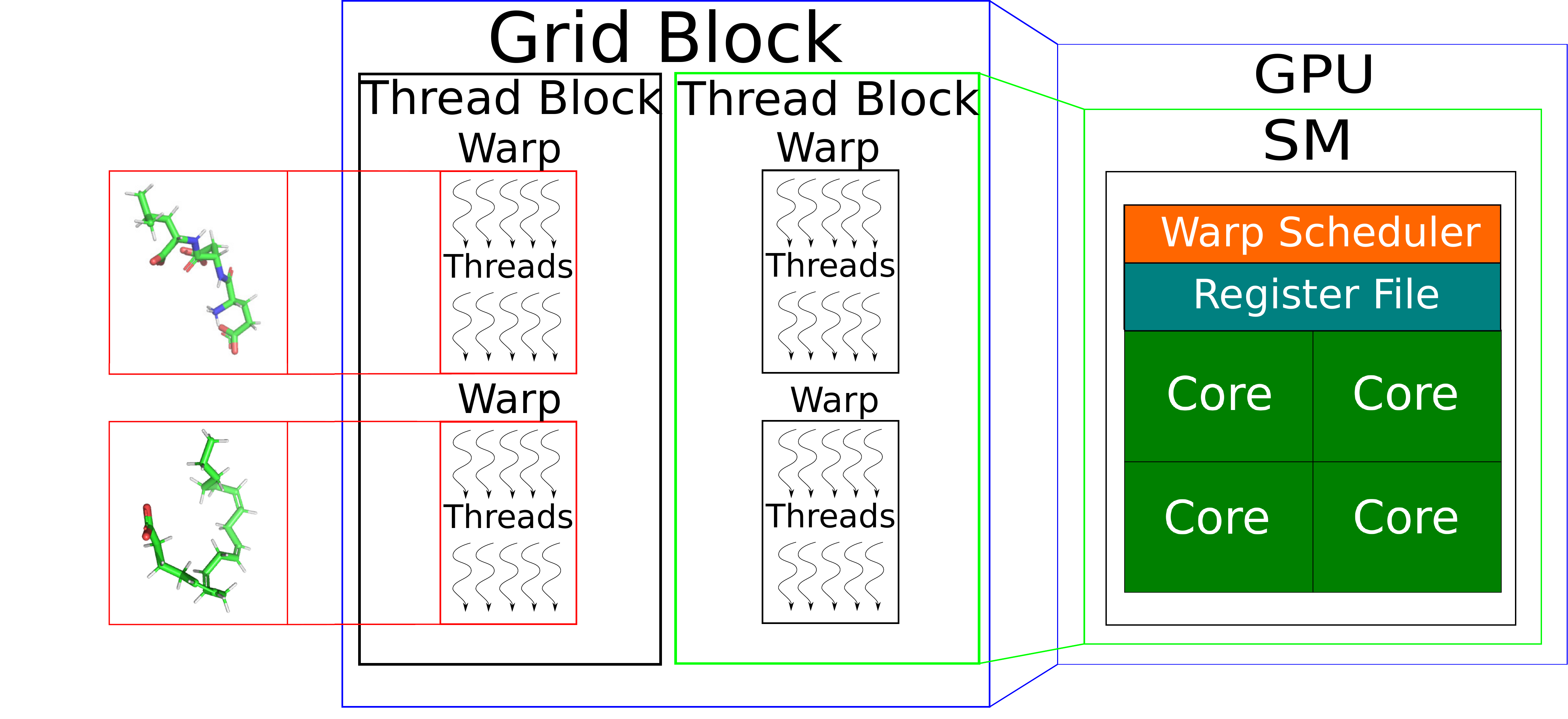}
    \caption{The mapping of ligands' computation on the GPU logical and physical architecture.}
    \label{fig:gpu-arch}
\end{figure}

\subsection{The \ligen{} implementation}
\label{ssec:ligen_implementation}

We can follow two main paths to hinge on GPU parallelism.
The most common one spreads the computation across the GPU to lower the execution time as much as possible.
AutoDock-GPU and \ligen{} developers followed that path \cite{10.1177/10943420211001565}  \cite{9817028} for an extreme-scale virtual screening campaign against SARS-CoV-2.
Since the computation of a single ligand is independent from the others, a second approach reduces its computation on a few GPU threads, to enable multiple ligand computations in parallel on the same device, as depicted in \prettyref{fig:gpu-arch}.
This approach increases the execution time to compute a single ligand compared to the CPU, but it can virtual screen a higher number of ligands in parallel bundled in a \textit{bucket}.
Literature already analyzed this approach \cite{10.48550/arxiv.2209.05069}, which provides a speed-up of up to $5$x w.r.t. the first one due to the lack of required synchronization.

From the implementation point of view, \ligen{} uses several CPU threads, named \textit{worker}s, to launch the computation kernels with a double buffering technique for hiding data transfers to and from the device.
In particular, when a worker receives a bucket of ligands, it starts to copy the data to the first available buffer on the GPU.
Then, it waits until the GPU is available and executes the code in mutual exclusion.
Finally, it copies the results on CPU memory and releases the buffer for another worker.
It is possible to use this approach for each available GPU in the system.

\section{Out-of-kernel optimizations}
\label{sec:ook_optimization}

The hardware requirements of a GPU kernel, such as registers and shared memory, strongly depend on the ligand's number of atoms.
The problem is that the kernels need to reserve memory for the worst case, severely hindering the performance since the atoms number can range from less than $20$ up to a few hundred.
We solve this issue with a different kernel for each ``class'' of ligands, using a non-type template parameter for the kernel maximum number of atoms and select the correct kernel version at runtime.
In this way we can instantiate a kernel tailored on a bucket of ligands with a similar number of atoms to minimize the resource waste.
Moreover, to further improve efficiency, we would like that the computations of all the ligands in a bucket have a similar execution time, i.e., they are balanced.
We can notice from the algorithm that its complexity is strongly related to the number of atoms and \textit{rotamers} \cite{9817028}.
The latter is a subset of the molecule atoms that ca    n rotate, changing its geometrical shape without altering physical and chemical properties.

Therefore, we need to create buckets of ligands with a similar number of atoms and rotamers.
For this reason, we introduce a runtime input preparation phase that collects the ligands in buckets before virtual screening them.
This approach raises two design choices: \textit{i)} how many ligands we want in each bucket, and \textit{ii)} how many clusters do we need.

To solve the first problem, \ligen{} queries the CUDA runtime, using the \texttt{cudaOccupancyMaxActiveBlocksPerMultiprocessor} function, to compute how many block $b$ can run on the same Streaming Multiprocessors (SM).
Then, we compute the number of ligands for each bucket $l$ on the fly as follows:
\begin{equation}
\label{eqn:cuda}
l = b \times SM \times \frac{t}{ws}
\end{equation}
where $SM$ is the number of available SM on the GPU, $t$ is the number of CUDA threads in a CUDA block, and $ws$ is the warp size.
Since we use one warp to process a ligand, $t$ is a tunable parameter that states how many ligands we want to compute in a CUDA block.
In the current version we virtual screen  one ligand per CUDA block.
We evaluate the impact of changing the number of ligands per bucket in \prettyref{sec:trace}

To address the second design choice, we consider the two clustering criteria as orthogonal since they have a weak relationship.
Thus, starting from an upper bound on the maximum number of atoms and rotamers, we need to define a heuristic to generate the intermediate values. 
Clusters generation uses atoms and warps sizes to balance computation among CUDA threads and hardware resources.
For example, if we would like to have three clusters for the atoms, we will bundle in the same buckets  ligands with up to $32$ and $64$ atoms while grouping the remaining ones in the last bucket.
For the rotamers, we use a heuristic that divides the range of values unevenly.
In particular, we are more interested in creating clusters toward lower values since they lead to the highest relative imbalance.
We evaluate the impact of this design parameter in \prettyref{ssec:heatmap}.
\section{SYCL porting}
\label{sec:sycl_porting}

The out-of-kernel optimization described in the previous section prepares the input to increase the computation efficiency on GPU.
While the methodology is independent of the kernel implementation, we use information from the CUDA runtime to compute how many ligands we want in a bucket.
The porting in SYCL 2020\footnote{\url{https://registry.khronos.org/SYCL/specs/sycl-2020/html/sycl-2020.html}} of this feature uses kernel bundles information from the standard.
Thus, we need to compute the number of ligands $l$ in a bucket as follows:
\begin{equation}
\label{eqn:sycl}
l = \frac{wgs}{t} \times CU \times \frac{t}{sgs}
\end{equation}
where $wgs$ is the maximum number of work items, i.e. CUDA threads, that we can execute on a single Compute Unit (CU), i.e. SM.
$CU$ is the number of CU, and $sgs$ is the maximum sub-group size, i.e. warp size.
Finally, $t$ is the work group size, i.e. the number of CUDA threads in a CUDA block.
Even if the latter can be simplified, we have written \prettyref{eqn:sycl} to match the terms in \prettyref{eqn:cuda}.
We can query $wgs$, $sgs$, and $CU$ using the SYCL 2020 API.
In particular,  $wgs$ is the property \texttt{kernel\_device\_specific::work\_group\_size}.
The parameter $t$ has the same value as in the CUDA implementation.
Unfortunately, if the SYCL compiler does not comply with the SYCL 2020 standard, the user needs to perform a manual tuning campaign.

\section{Experimental results}
\label{sec:experimental_results}

In this section, we report the experimental campaign results on the effect of the out-of-kernel optimizations.

\subsection{Experimental setup}

All the experiments use a single node from the Karolina\footnote{\url{https://www.it4i.cz/en/infrastructure/karolina}} supercomputer at IT4I.
It has two AMD EPYC 7763 @ $2.53$GHz and eight NVIDIA A100 GPU.
We compile \ligen{} $5.0$ with the support for both kernel implementations.
We use the Intel(R) oneAPI DPC++/C++ Compiler $2023.0.0$ to compile C++ and SYCL sources, while we use NVCC $11.7.99$ to compile the CUDA source code.

We use the average application throughput to measure the application performance, since this is a typical HPC batch job.
To measure the throughput we divide the number of processed molecules by the elapsed walltime.

\subsection{Optimizing the bucket size}
\label{sec:trace}

\begin{figure}[t]
    \centering
    \begin{subfigure}[b]{0.47\textwidth}
        \centering
        \includegraphics[width=\textwidth]{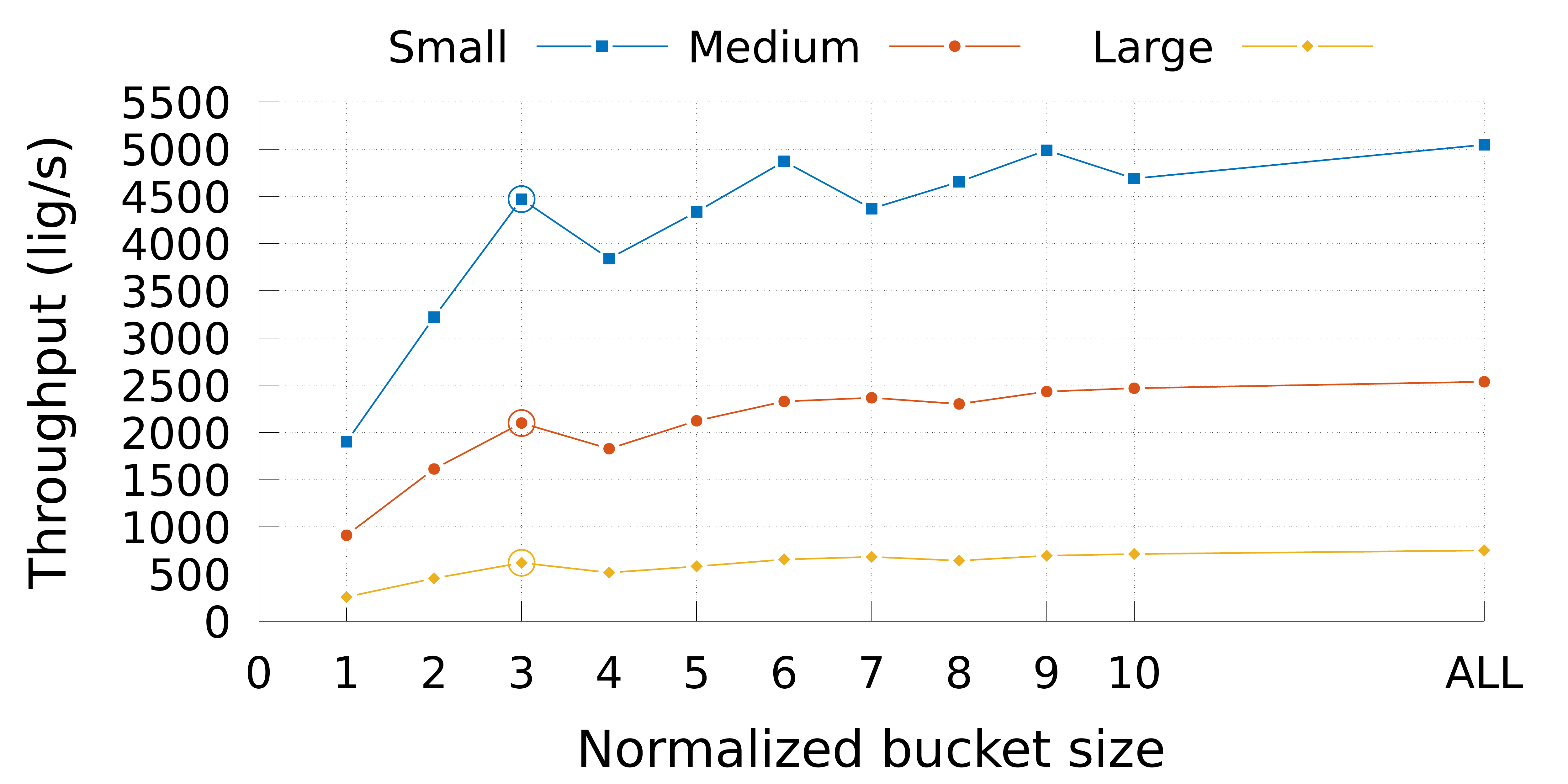}
        \caption{CUDA implementation}
        \label{fig:throughputPlotCuda}
    \end{subfigure}
    \hfill
    \begin{subfigure}[b]{0.47\textwidth}
        \centering
        \includegraphics[width=\textwidth]{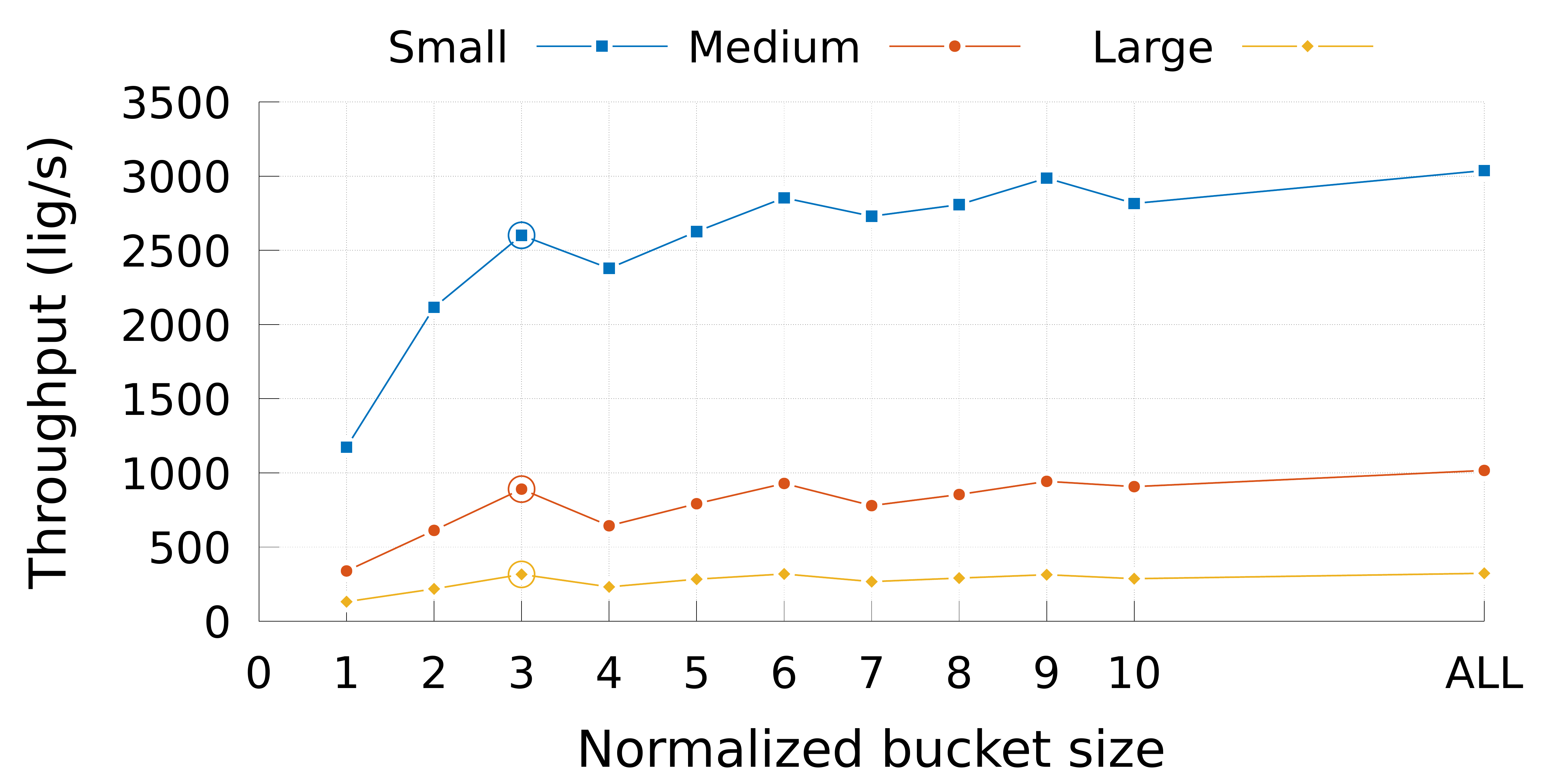}
        \caption{SYCL implementation}
        \label{fig:throughputPlotSycl}
    \end{subfigure}
    \caption{\ligen{} average throughput by varying the number of ligands in a bucket, for three classes of ligands.}
    \label{fig:throughputTrace}
\end{figure}

\begin{figure}[t]
    \includegraphics[width=0.47\textwidth]{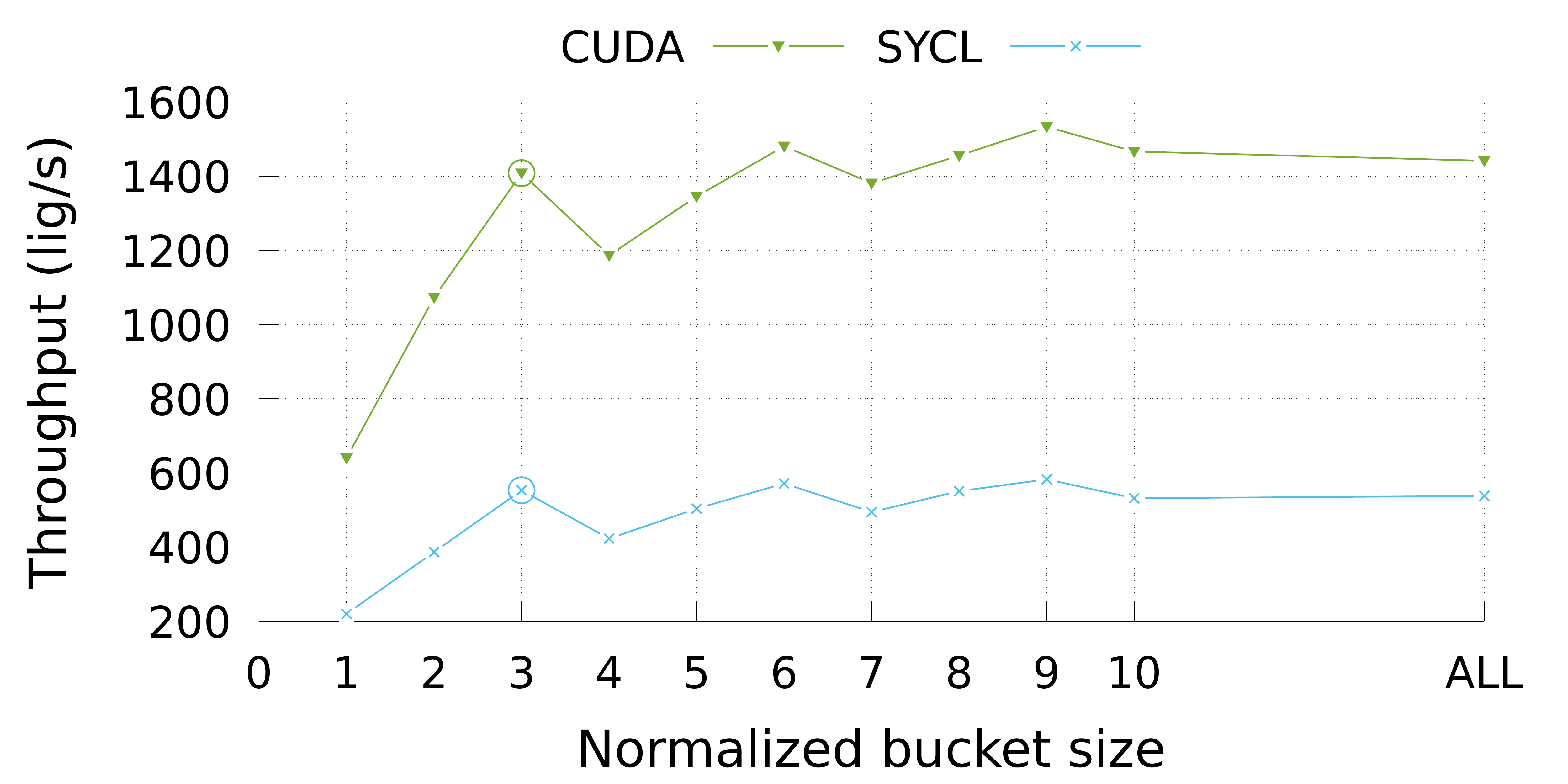}
    \caption{\ligen{} average throughput with a mixed dataset of ligands, by varying the number of clusters.}    \label{fig:throughputMixedPlot}
\end{figure}

In this experiment we measure the computation efficiency according to the number of ligands in a bucket.
To reduce the impact of different features, such as the number of rotamers, we focus on three ligands: ``small'', ``medium'', and ``large''.
They have respectively $17$, $53$, and $99$ atoms.
We measured a stable throughput by replicating each ligand $10$ million times and docked the ligands in twelve docking sites of a target protein.

\prettyref{fig:throughputTrace} shows the experiment results using the CUDA (\prettyref{fig:throughputPlotCuda}) and SYCL (\prettyref{fig:throughputPlotSycl}) implementations.
The x-axis represents the number of ligands in each bucket.
We normalized the number of ligands in a bucket by a third of the value suggested by \prettyref{eqn:cuda} or \prettyref{eqn:sycl}, as reported in \prettyref{tab:comparison}.
We expect to maximize the GPU utilization every three points.
Moreover, we evaluate the application performance using the maximum number of ligands that fits in the GPU memory, with the label ``All''.

We can notice two phases from the experimental results: before and after the value suggested by the equations.
Before this value, the application throughput increases with the number of ligands.
This is due to the increasing number of active warps that the SM scheduler can use to hide computation stalls.
After this value, we have a drop in the performance since we are processing the same ligand, and its computation will last for a similar amount of time.
Then, the throughput will increase as more active warps are created, repeating the pattern.
The pattern is more evident for the SYCL and CUDA implementation with small ligands. While for medium and large ligands the change in throughput is more smooth.

The throughput trend suggests using a multiple of the number of ligands equal to the value computed with \prettyref{eqn:cuda} or \prettyref{eqn:sycl}.
%
Ideally, we would like to use the highest multiple that the GPU memory allows.
However, the ligands' memory footprint grows linearly with the number of docking sites founds in the virtual screening campaign.
For this reason, a high number of ligands per bucket will limit the number of docking sites that we can consider according to the capacity of the target GPU.

To measure the performance in a more realistic scenario, we use the heterogeneous dataset reported in \prettyref{ssec:heatmap}.
\prettyref{fig:throughputMixedPlot} show the performance of the application for the CUDA and SYCL implementation.
To reach this goal we use $6$ clusters for the number of atoms and $23$ clusters for the rotamers.
The experimental results confirm the execution patter, suggesting that the number of ligands computed with \prettyref{eqn:cuda} and \prettyref{eqn:sycl} with different input features is reliable.

\begin{table}[b]
\centering
\footnotesize
\caption{\label{tab:comparison}Comparison between the CUDA and SYCL kernel implementations, by varying the maximum number of atoms.}
\begin{tabular}{@{}|r|cc|cc|@{}}
\hline
\multicolumn{1}{|c|}{\multirow{2}{*}{\textbf{Num atoms}}} & \multicolumn{2}{c|}{\textbf{CUDA}}                     & \multicolumn{2}{c|}{\textbf{SYCL}}                     \\ \cline{2-5}
\multicolumn{1}{|c|}{}                           & \multicolumn{1}{c|}{Register} & Active blocks & \multicolumn{1}{c|}{Register} & Active blocks \\ \cline{2-5}
\hline
32                                                & \multicolumn{1}{c|}{104}        & 1728             & \multicolumn{1}{c|}{160}        & 1296             \\
64                                                & \multicolumn{1}{c|}{91}        & 2160             & \multicolumn{1}{c|}{174}        & 864             \\
96                                                & \multicolumn{1}{c|}{124}        & 1728             & \multicolumn{1}{c|}{178}        & 864   \\
\hline
\end{tabular}
\end{table}

When we compare the CUDA and SYCL performance, we analyzed the most compute intensive kernel that count for more than $90\%$ of the total execution time, i.e. the docking kernel, since we noticed a larger slowdown than expected.
%
Indeed, the execution time of the main kernel in SYCL is $12\%$ slower than CUDA, while the gap in the application throughput is wider.
%
This is due to a higher register pressure of the SYCL implementation, as reported in \prettyref{tab:comparison}, which reduces the parallelism achievable between kernels.
So that SYCL can process a smaller number of ligands in parallel.
%
The efficiency of using the eight GPUs available on the node is the same across the two implementations

\subsection{Measuring the computation efficiency}
\label{ssec:heatmap}


\begin{figure}[t]
    \centering
    \begin{subfigure}[b]{0.47\textwidth}
        \centering
        \includegraphics[width=\textwidth]{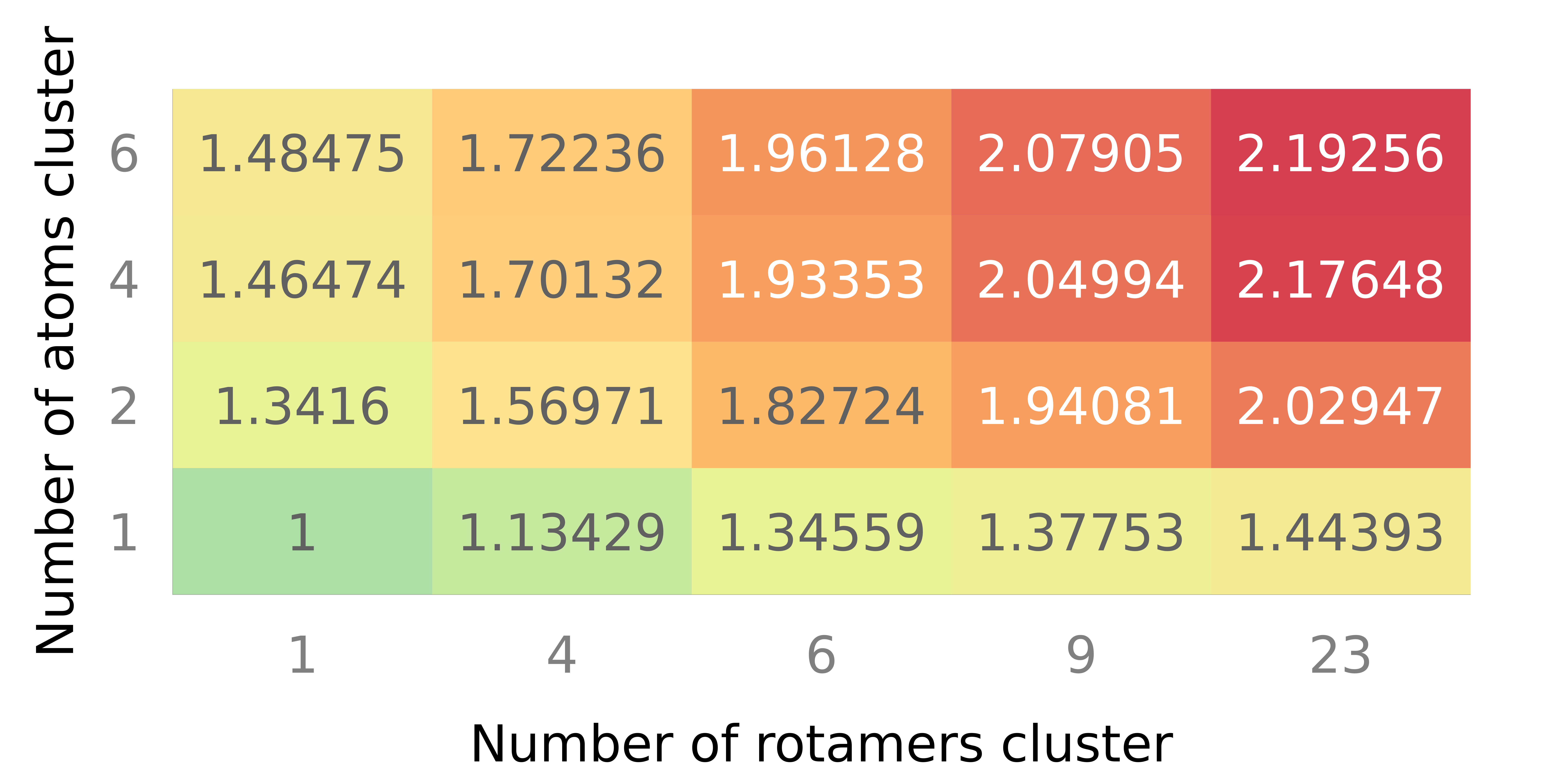}
        \caption{CUDA implementation}
        \label{fig:heatmap-cuda}
    \end{subfigure}
    \hfill
    \begin{subfigure}[b]{0.47\textwidth}
        \centering
        \includegraphics[width=\textwidth]{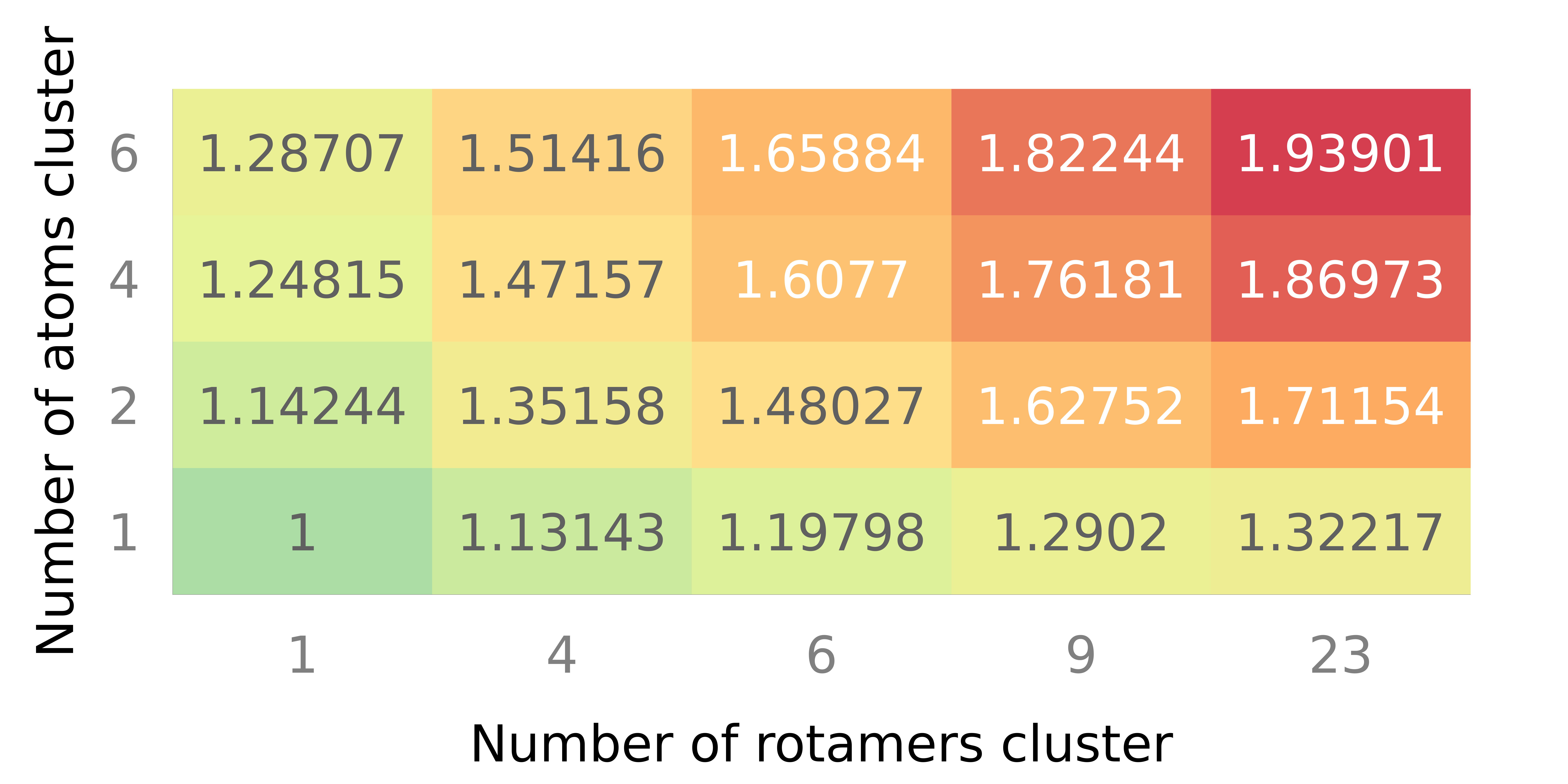}
        \caption{SYCL implementation}
        \label{fig:heatmap-sycl}
    \end{subfigure}
    \caption{Application throughput speed-up by varying the number of clusters that targets both, the number of atoms and rotamers in a ligand.}
    \label{fig:heatmap}
\end{figure}

This experiment aims at measuring the impact on the performance of the clustering procedure. We reach this goal with a virtual screening campaign on a dataset of heterogeneous ligands. The input dataset contains $10$ million different ligands with an atoms numbers that range between $20$ and $120$, while they have from $0$ to $20$ rotamers.
We use \prettyref{eqn:cuda} and \prettyref{eqn:sycl} to set each bucket size.
\prettyref{fig:heatmap} shows the speed-up of the application throughput by varying the number of clusters for rotamers (x-axis) and the atoms (y-axis).
We use as baseline a single cluster for both characteristics.

The experimental results show how it is possible to roughly double the application throughput by using the analyzed out-of-kernel optimization.
Also, the trend along the x- and y-axis suggest that we improve the computation efficiency when the ligands in a bucket share similar features.
Moreover, the benefits of this optimization are consistent across the two kernel implementations.

However, when we increase the number of clusters, we spread the input dataset across more buckets.
Therefore the number of half-empty buckets will be high at the end of the execution, decreasing the computation efficiency.
Therefore, we need many ligands' to benefit from a fine-grained clusterization.
This constraint is not problematic for extreme-scale virtual screening campaigns, but is relevant for day-by-day usage.

\section{Conclusion}
\label{sec:conclusion}

In this paper, we explored the impact of an input data preparation phase that use input and architecture features to improve the computation efficiency on GPUs.
We used a real-world virtual screening application as a case study to evaluate their impact and limitations.
However, is possible to use the same approach with applications that share a similar design.

From the experimental results, we can notice how it is possible to improve the computation efficiency, almost doubling the application throughput.
This gain in performance is consistent with both the SYCL and CUDA implementation.
However, a large or homogeneous input set is required to reach this efficiency level.
This is not a problem since we need the maximum efficiency only on extreme-scale virtual screening campaign.


\begin{acks}
This section is omitted for the double blind review.
\end{acks}

\bibliographystyle{ACM-Reference-Format}
\bibliography{main}

\end{document}